# Manipulating composition gradient in cuprate superconducting thin films

Heshan Yu[1,2], Jie Yuan[1,3] *, Beiyi Zhu[1], & Kui Jin[1,2,3,4] *

[1]*Beijing National Laboratory for Condensed Matter Physics, Institute of Physics, Chinese Academy of Sciences, Beijing 100190, China*
[2]*University of Chinese Academy of Sciences, Beijing 100049, China*
[3]*CAS Key Laboratory of Vacuum Physics, University of Chinese Academy of Sciences, Beijing 100049, China*
[4]*Collaborative Innovation Center of Quantum Matter, Beijing, 100190, China*

The techniques of growing films with different parameters in single process make it possible to build up a sample library promptly. In this work, with a precisely controlled moving mask, we synthetized superconducting $La_{2-x}Ce_xCuO_{4\pm\delta}$ combinatorial films on one $SrTiO_3$ substrate with the doping levels from $x$ = 0.1 to 0.19. The monotonicity in doping along the designed direction is verified by micro-region x-ray diffraction and electric transport measurements. More importantly, by means of numerical simulation, the real change of doping levels is in accordance with a linear gradient variation of doping levels in the $La_{2-x}Ce_xCuO_{4\pm\delta}$ combinatorial films. Our results indicate that it is promising to accurately investigate materials with critical composition by combinatorial film technique.

The high-throughput strategy has been widely developed in bioinformatics and pharmaceutical industry [1]. The combinatorial method is the most crucial experimental technique to put this strategy into practical application [2]. The main idea is synthesizing samples with various physical or chemical parameters in one process and characterizing them simultaneously. In the 1960s, this idea had been introduced to the material research by Hanak [3]. Unfortunately, owing to the lack of powerful methods used for data collection and analysis, this forward-looking attempt was ignored.

In the past two decades, the developments in related technology, especially in information technology have greatly paved the way for the combinatorial experiments. In 1995, Xiang *et al.* successfully fabricated a 128-member library of copper oxide thin films on a single substrate [4]. Since then the combinatorial methods have been widely used to screen luminescent materials, dielectric materials, multiferroic materials, ferromagnetic shape memory alloys, and multifarious other materials [5-14], which greatly raise the efficiency of material synthesis and screening.

Among the combinatorial methods, the continuous composition spread film (CCSF) technique could provide one- or two-dimensional spread of the doping compositions. The basic idea of CCSF synthesis is to realize different continuous distributions of a given binary or ternary system on one substrate and to make precusors combined with each other[15-18]. Then continuous chemical composition could be achieved in one piece of film, i.e. combi-film. The evolution of physical properties versus composition could be mapped out via characterizing the individual small regions of the film by patterning the sample into small chips or submitting it to a facility with high spatial resolution.

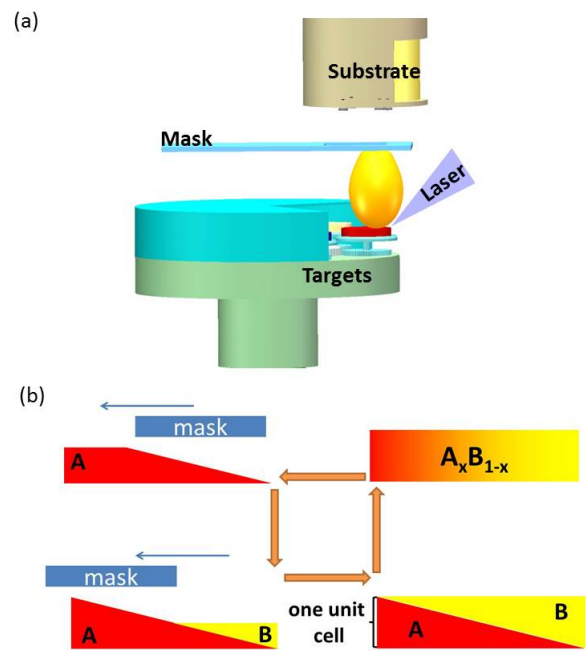

Figure 1 (Color online) (a) The schematic diagram of the continuous moving mask technique; (b) One typical procedure of binary combi-film growth by using moving mask technique.

In 2000, Fukumuri *et al.* used the continuous moving mask technique to make the composition distribution become well controlled [19]. Figure 1(a) shows the schematic diagram of the continuous moving mask technique. For clarity, a schematic procedure for a binary system is illustrated in Figure 1(b). When target A is ablated by laser pulses, a metal mask moves at a constant speed and passes through over the substrate gradually, which is expected to cause a linear distribution of precursor A on the substrate in half a period. Then target B is switched to face the substrate and the mask moves in the opposite direction to get reversal distribution of precursor B in the other half period. Such a process will be repeated periodically to reach an aimed thickness. The parameters, such as deposition rate, laser pulses in one period, moving speed of the mask, etc., are strictly controlled to avoid superlattice structure.

*Jie Yuan, email: yuanjie@iphy.ac.cn;
*Kui Jin, email: kujin@iphy.ac.cn

We attempt to fabricate combinatorial $La_{2-x}Ce_xCuO_{4\pm\delta}$ ($x$ = 0.1 ~ 0.19) thin films on one $SrTiO_3$ (STO) substrate with a programmed-motion mask. The targets A and B are $La_{2-x}Ce_xCuO_{4\pm\delta}$ ($x$ = 0.1) and $La_{2-x}Ce_xCuO_{4\pm\delta}$ ($x$ = 0.19), respectively. After eighty periods, a 100 nm thick combinatorial $La_{2-x}Ce_xCuO_{4\pm\delta}$ (LCCO) film is achieved after the in-situ reduction process of several minutes in vacuum at about 700℃.

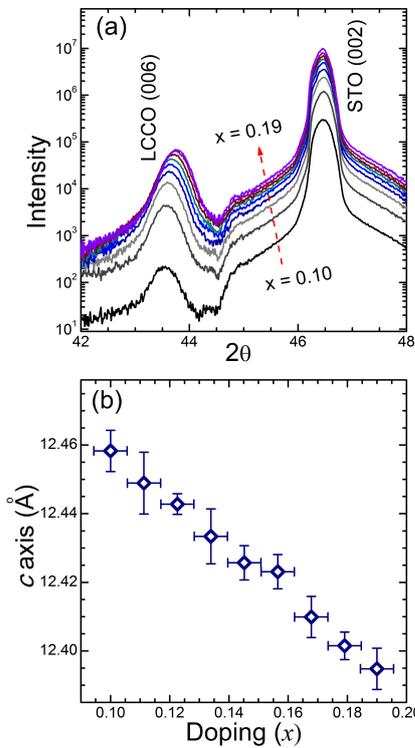

Figure 2 (Color online) (a) The micro-region X-ray diffraction results of combi-film $La_{2-x}Ce_xCuO_{4\pm\delta}$. The component interval $\Delta x$ is about 0.012. (b) The variation of $c$-axis lattice constant with increasing doping levels. In this figure, $x$ refers to nominal doping level.

The crystalline of the whole film is tested by an x-ray diffract meter with the micro area measurement unit (Rigaku, SmartLab9KW). The beam size is 0.4 mm in width and set to scan the film from one end to the other. In every step, the $2\theta$ is taken from 10° to 80° to characterize the structure of a small region in the combi-film. In Figure 2(a), the micro-region x-ray diffraction data demonstrates the combinatorial thin film grows with single orientation (00$l$). With moving the x-ray spot to the high Ce-doping side, the LCCO (006) diffraction peak gradually moves to higher angle whereas the position of STO (002) peak is fixed. The lattice constant of $c$-axis can be calculated by the Bragg's formula $2dsin\theta = n\lambda$, where $\theta$ is the diffraction angle. As shown in Figure 2(b), the lattice constant of $c$-axis monotonically decreases with raising the nominal doping, in accordance with the evolution of the $c$-axis lattice constant for single component LCCO films fabricated by the conventional pulsed laser deposition (PLD) technique [20, 21].

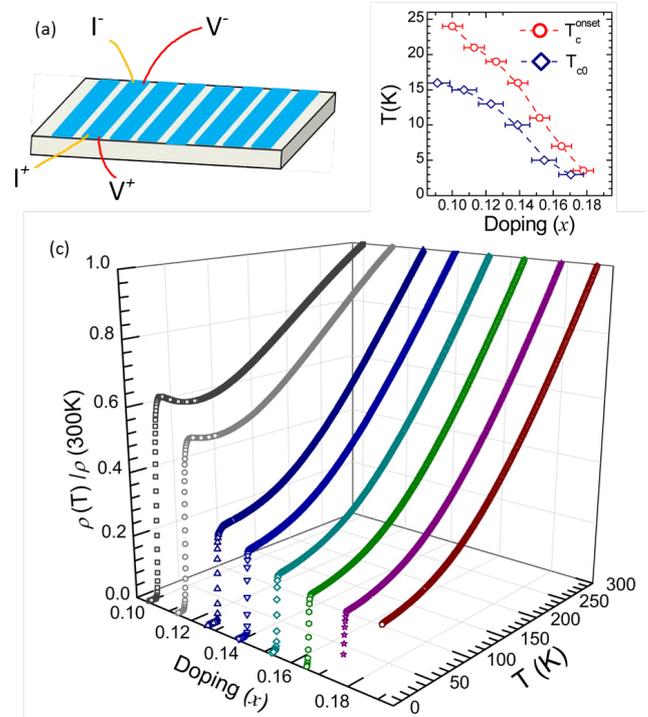

Figure 3 (Color online) (a) The schematic diagram of patterns in electric transport measurements. (b) The variation of zero-resistance transition temperature $T_{c0}$ and onset of superconducting transition $T_c^{onset}$ with increasing nominal Ce doping levels. (c) The 'resistivity versus temperature' curves at different doping levels. In this figure, $x$ refers to nominal doping level.

Then the combinatorial thin films are patterned into small chips (see Fig. 3a), to measure electrical resistivity from 300 K to 2 K. As shown in Figure 3(c), all the channels show metallic behaviors above 50 K. With further reducing the temperature, a superconductor-to-metal transition occurs between $x$ = 0.17 and 0.19, that is, superconductivity fades away with approaching $x$ = 0.19. At $x$ = 0.19, the resistivity shows $T^2$-behavior. A lightly upturn is seen in the $\rho$-$T$ curves at low temperature close to the nominal optimal doping level $x$ = 0.10, implying that there is a minute amount of excess oxygen in the combi film. In Figure 3(b), both the zero-resistance transition temperature $T_{c0}$ and the onset of superconducting temperature $T_c^{onset}$ decrease with increasing Ce doping, in accordance with previous studies [21, 22]. The structural and electrical transport characterizations demonstrate that the combinatorial LCCO films have been successfully fabricated by the continuous moving mask technique.

Through the x-ray and transport measurements, a chemical composition gradient is verified to consist in the combinatorial LCCO films. It should be noted that the linear relation between Ce doping level and spatial location in the film is based on the assumption of a constant deposition rate.

In fact, each plume can persist in a scale of microsecond (μs) as revealed by the light emission experiments, that is, the matter deposited on the substrate in each plume is pulsed [23]. The speed of the mask is ~0.6 mm/s and for each step the mask moves several μm, which means the time resolution of each step is in order of magnitude of millisecond (ms), which is much longer than the duration of matter deposition. Thus, the discontinuity is restricted by the pulsed matter deposition instead of the resolution of mask motion in our system. Screening a micro-region larger than several μm, one does not have to consider the discontinuity. However, for the high spatial resolution instruments, such as the scanning tunneling microscope down to the level of atom resolution, the discontinuity should be carefully treated. In the following section, we will simulate the growth process of our combi-film to estimate how the real composition gradient deviates from the ideal design.

In this simulation, the selected model is usually used in PLD [24, 25]. Meanwhile, the mask motion and shadowing effects are taken into account [26]. Without shadowing, average and diffusion effects, etc., the film deposited on a substrate will have steep edges. Then the film grown by PLD with a mask moving in one direction will have ladder-like edges as shown in Figures 4(a) and (b), since the composition of the combi-film is mainly determined by the distribution of the precursors. In this manner, the largest deviation from the ideal linear composition gradient should be located at the edges of each step, which is equal to $\Delta x = (x_2 - x_1)/N$, where $x_1$ and $x_2$ stand for micro-region compositions within the film, $N$ is the number of laser pulses for one target in one period. It seems that one can improve the resolution to approach a real linear distribution by increasing $N$. However, the real deposition procedure always involves the shadowing effect and gives rise to a tail at the edges as shown in Figure 4(c).

The shape of the tail will be affected by the distance between mask and substrate, and the atmosphere during the deposition, etc. We assume that the shape of the tail obeys the relation $f \sim e^{-l/\lambda}$ [Figure 4(d)], where $f$ is the equivalent film thickness, $l$ is the distance from the ideal edge and $\lambda$ is the effective diffusion length. Then we simulate the growth procedure using laser pulses $N = 100$, and divide the substrate into one dimensional uniform grid with an interval of 10 μm, along the mask moving direction. Then, the position dependence of the chemical composition can be achieved by calculating the ratio of two precursors deposited at each node.

Firstly, we tune $\lambda$ from 1 to 1000, and get one thousand curves of Ce($x$) vs. position, ten of which are shown in Figure 4(e). The $\Delta x$ reduces as $\lambda$ grows. In the inset of Figure 4(e), the terraces by pulsed deposition are smeared out with increasing $\lambda$.

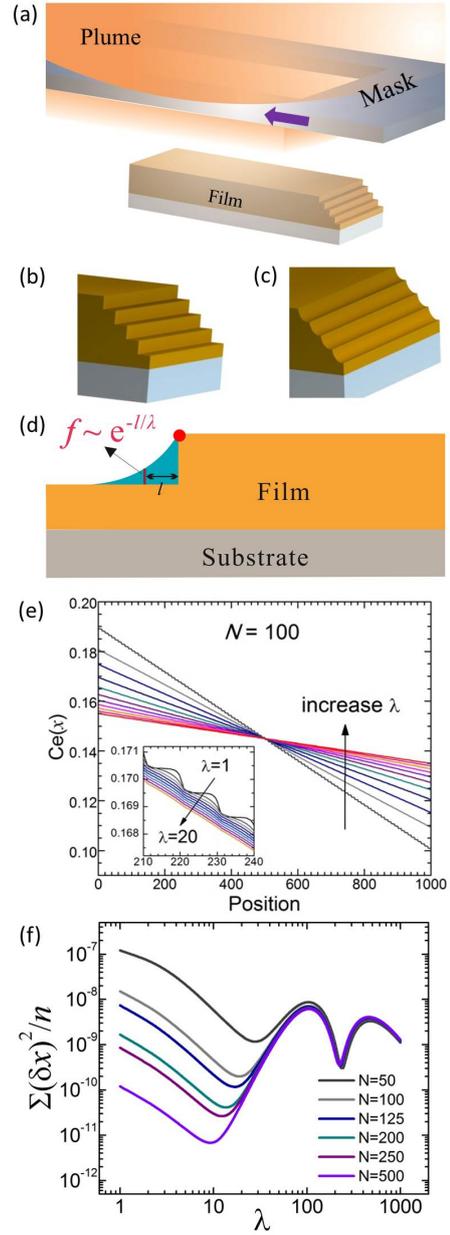

Figure 4 (Color online) (a) The schematic diagram of the combi-film growth regardless of shadowing effect. (b) The schematic diagrams of the film morphology without shadowing effect. (c) The schematic diagrams of the film morphology with shadowing effect. (d) The schematic diagram of a tail. (e) The simulated curves of $x$ vs. position with different $\lambda$. The inset is the zoom-in view. (f) The simulated curves of mean squared error $\Sigma \delta x_i^2 / n$ vs. $\lambda$ with different laser pulses N.

Secondly, depending on the linear least square method $\sigma = \Sigma \delta x_i^2 / n$, we fit the data of $x$ vs. position to a straight line, where $\delta x_i$ is the composition deviation from the fitting line at each position and $n$ is the number of positions in the simulations. The mean squared error for each $\lambda$ is shown in Fig.4(f). With increasing $N$, $\sigma$ decreases significantly for a smaller $\lambda$ but keeps nearly the same for a higher $\lambda$, which indicates that $\delta x_i$ is controllable by tuning the growth pa-rameters. That is, one can approach an ideal linear distribution of the compositional gradient by increasing the number of laser pulses in one period for a smaller λ. Meanwhile, by varying the deposition pressure or the distance between the mask and the substrate the diffusion length can be tuned.

For instance, one picks $\Delta x$ = 0.085, 0.08, 0.075, 0.07, 0.065 and 0.06, then gets the corresponding $\sigma = 2.17 \times 10^{-10}$, $2.37 \times 10^{-9}$, $6.32 \times 10^{-9}$, $6.74 \times 10^{-9}$, $3.71 \times 10^{-9}$ and $9.21 \times 10^{-9}$. Considering $\Delta x$ = 0.06, $\delta x = 2.60 \times 10^{-4}$, the actual deviation at each point will be less than $2.60 \times 10^{-4}$. This value is much better than the controllability in synthesizing LCCO thin films, which is $\sim 10^{-3}$ [27].

To summarize, we employ the continuous moving mask technique to fabricate LCCO combi-films with nominal doping levels from $x$ = 0.10 to $x$ = 0.19. The micro-region x-ray diffraction reveals that the films are of high quality and in single orientation. Both the zero-resistance transition temperature $T_{c0}$ and lattice constant of $c$-axis monotonically decrease with increasing Ce, consistent with those of uniform samples fabricated by the conventional PLD technique. According to the numerical simulation involving shadowing and diffusion effects, the deviation from desired linear distribution of chemical composition is estimated to be of the order of $\sim 10^{-4}$, better than the accuracy of traditional synthesis methods ($\sim 10^{-3}$). In other words, the success in synthesizing such combi-films will promote the research which requires precise controlment on doping level, i.e. quantum phase transition as well as the quantum critical points which require the deviation of composition should be smaller than $10^{-3}$ [22, 28-30]. Consequently, the CCSF technique will make a breakthrough for mapping a more precise phase diagram [31].

*This work is supported by the Key Research Program of Frontier Sciences, CAS, Grant (QYZDB-SSW-SLH008), National Natural Science Foundation of China (11474338, 11674374), the National Key Basic Research Program of China (2015CB921000, 2016YFA0300301).*